# A Novel Real-Time Energy Management Strategy for Gird-Friendly Microgrid: Harnessing Internal Fluctuation Internally

Cunzhi Zhao
*Student Member, IEEE*
Department of Electrical and Computer Engineering
University of Houston
Houston, TX, USA
czhao20@uh.edu

Xingpeng Li
*Member, IEEE*
Department of Electrical and Computer Engineering
University of Houston
Houston, TX, USA
xli82@uh.edu

*Abstract--* Typically, a large portion of microgrid generating capacity is from variable renewable resources that are greatly impacted by the environment and can be intermittent as well as stochastic. This would result in uncertainty of microgrid net-load, and negatively affect the grid reliability. A two-phase real-time energy management strategy for networked microgrid is proposed in this paper to address microgrid internal fluctuation internally, which enables a microgrid to become grid-friendly. The proposed strategy is based on coordination between the real-time dispatch (RTD) phase and the real-time control (RTC) phase. In the RTD phase, model predictive control (MPC) is used to optimally dispatch microgrid resources in the current time interval while considering near future situations. The RTC phase addresses microgrid internal net-load fluctuation with fast-acting batteries, which aims to maintain a constant tie-line power flow between the main grid and the microgrid for the current dispatch interval. Numerical simulations conducted on ten different net-load scenarios can demonstrate the performance of the proposed two-phase energy management strategy that will enable a microgrid to operate as a controllable asset with static electricity consumption or production in an economic dispatch interval from the perspective of the bulk grid operator.

*Index Terms—* Battery energy storage system, Controllable microgrid, Distributed energy resources, Grid-friendly microgrid, Microgrid energy management, Model predictive control, Multi-time-scale coordination.

## Nomenclature

| | |
|---|---|
| $g$ | Generator index. |
| $t$ | Time period index. |
| $G$ | Set of generators. |
| $I$ | Set of uncontrollable generators. |
| $E$ | Set of energy storage systems. |
| $c_g$ | Linear cost for generator $g$. |
| $c_g^{NL}$ | No load cost for generator $g$. |
| $c_g^{on}$ | Start-up cost for generator $g$. |
| $\Delta T$ | Length of a single dispatch interval. |
| $\Delta T_c$ | Length of a single control interval. |
| $R_{percent}$ | Percentage of the backup power to the total power. |
| $E_s^{max}$ | Maximum energy capacity of ESS $s$. |
| $E_s^{min}$ | Minimum energy capacity of ESS $s$. |
| $U_{grid+}^t$ | Status of purchasing power from main grid in time period $t$. |
| $U_{grid-}^t$ | Status of selling power to main grid status in time period $t$. |
| $c_{grid+}^t$ | Wholesale electricity purchase price in time period $t$. |
| $c_{grid-}^t$ | Wholesale electricity sell price in time period $t$. |
| $U_{s+}^t$ | Charging status of energy storage system $s$ determined in RTD phase in time period $t$. It is 1 if charging status; otherwise 0. |
| $U_{s-}^t$ | Discharging status of energy storage system $s$ determined in RTD in time period $t$. It is 1 if discharging status; otherwise 0. |
| $P_{grid+}^t$ | Purchase from main grid power in time period $t$. |
| $P_{grid-}^t$ | Sell to main grid power in time period $t$. |
| $P_{grid}^c$ | Tie-line exchange power in a 4-second RTC interval $c$. |
| $P_{gt}$ | Output of generator g in time period $t$. |
| $P_I^t$ | Output of uncontrollable power unit at time period t. |
| $P_L^t$ | Internal demand of microgrid at time period t. |
| $P_{st-}$ | Discharging power of energy storage system $s$ in time period $t$. |
| $P_{st+}$ | Charging power of energy storage system $s$ in time period $t$. |
| $P_{grid}^{max}$ | Maximum thermal limit of tie-line between main grid and microgrid. |
| $P_g^{max}$ | Maximum capacity of generator $g$. |
| $P_g^{min}$ | Minimum capacity of generator $g$. |
| $P_g^{Ramp}$ | Ramping limit of generator $g$ |
| $P_S^{max}$ | Maximum charge/discharge power of ESS $s$. |
| $P_S^{min}$ | Minimum charge/discharge power of ESS $s$. |
| $\Delta P_s$ | Charging-discharging power capacity withhold for real-time control phase. |
| $\Delta E_s$ | Energy storage capacity withhold for real time-control phase |

## I. Introduction

Microgrid is a local asset aggregator that coordinates and manages distributed energy resources (DERs) in an autonomous and decentralized manner. A networked microgrid can operate (i) in a grid-connected mode with the main grid, or (ii) in an isolated mode without the main grid. In the grid-connected mode, microgrid remains connected to the main grid while importing or exporting power from/to the main grid. When there is a disturbance in the main grid that affects microgrid reliability, microgrid can switch to the isolated mode which can supply the power by itself [2]. DERs include energy storage system (ESS) and renewable energy resources (RESs) such as solar photovoltaic (PV) and wind turbines (WT) [3]–[4]. Renewable energy resources develop rapidly nowadays [4]. The real-time generation of those variable renewable units depends on weather situations such as solar irradiation and wind speed and can be highly intermittent and stochastic [5]. This leads to uncertainties in addition to the load fluctuation in the microgrid, thereby creating serious challenges for microgrid energy management [6]. With deployment of a large number of networked microgrids with high penetration of distributed RESs, the power grid will encounter unprecedented uncertainty spread over the entire systems.

Microgrid energy management is very important and it has been extensively studied in the literature. A microgrid load management is introduced in [7] to maintain the balance between generation and load; loads are classified based on whether the loads can be shed or not. However, only the near real-time case has been discussed in [7]. A chance constrained approach is proposed to systematically incorporate the energy management problem with uncertainties caused by RESs in the grid-connected microgrid [8]. Deep recurrent neural network leaning is introduced in [9] to minimize the microgrid cost without using information of long-term forecasting. A rolling horizon strategy that only covers a single interval is proposed here in [10] for RES based microgrid. The cost of microgrid is decreased by updating the optimized set points for DERs. Though [7]–[10] propose several effective microgrid energy management strategies, they only cover a single time interval in their real-time economic dispatch optimization. Single interval dispatch strategies fail to consider the variabilities and uncertainties associated with the very next few intervals which may affect the cost and reliability. In addition, those strategies ignore real-time sub-minute net-load fluctuation and assume the main grid can absorb the fluctuation at no cost. However, this could lead to substantial power grid uncertainty and may negatively affect system reliability significantly. One effective solution to this challenge is to develop a novel microgrid energy management strategy to harness the internal net-load fluctuation within the microgrid.

To address the aforementioned gaps, this paper proposes a two-phase real-time energy management strategy for grid-friendly microgrid. The proposed strategy can address the microgrid internal net-load fluctuation internally within the microgrid and enable the microgrid to perform as a controllable flat load during a dispatch interval. The proposed strategy consists of a real-time dispatch (RTD) phase and a real-time control (RTS) phase. In the RTD phase, instead of covering a single time interval, a rolling horizon based model predictive control (MPC) method that covers multiple intervals is applied to reduce the risk of dispatch failure and increase the reliability. In the RTC phase, the fast-acting battery energy storage system plays an important role of mitigating the net-load fluctuation and maintaining constant exchange power flow on the tie-line. As a result, the main grid will not be affected by the uncertainties within microgrids.

The rest of the paper is organized as follows. The proposed mathematical model is presented in Section II. Section III briefly discusses the details of the two-phase real-time energy management strategy. Case studies and discussion are presented in Section IV. Section V concludes the paper.

## II. MATHEMATICAL MODEL

### A. Real Time Dispatch

The RTD phase is to solve a multi-interval microgrid economic dispatch problem. Each time interval $\Delta T$ is 15 minutes. Its objective is to minimize the operation cost over the very next multiple dispatch intervals, which is shown below:

$$f(cost) = f_G + f_{grid} \qquad (1)$$

where $f_G$ denotes the cost of controllable DER units and $f_{grid}$ denotes the grid electricity cost. $f_G$ includes variable operation costs, as defined in (2). The cost of electricity trading with grid $f_{grid}$, as defined in (3), involves activities of buying power from the grid and selling power back to the grid. The purchasing and selling power prices are different. If the cost $f_{grid}$ is negative for a time period, it means the microgrid sells power to the grid and makes profit in that period.

$$f_G = \sum_{t \in S_T, g \in S_G}(c_g P_{gt}) \qquad (2)$$
$$f_{grid} = \sum_{t \in S_T}(c^t_{grid+} P^t_{grid+} - c^t_{grid-} P^t_{grid-}) \qquad (3)$$

Microgrid power balance equation is included in (4). Equation (5) calculates the ESS energy level. Constraint (6) ensures the microgrid to have sufficient backup power capacity for emergency. The maximum and minimum limits of controllable DERs are considered in (7). Also, the ramping limits of DERs are imposed in (8). The constraints of ESS charging or discharging rate limits are given in (9)–(10). Constraint (11) ensures that the energy storage system would not charge and discharge power at the same time. Constraint (12) ensures the energy stored in ESS is within maximum and minimum energy limits. As shown in (9)–(10) and (12), some capacity of ESS charging/discharging power and energy stored in ESS are withheld in the RTD phase for potential net-load fluctuation mitigation in the RTC phase. These reserved capacity would be deployed in the RTC phase. Tie-line exchange power limits are enforced in (13) – (14); constraint (15) ensures that the microgrid would not purchase and sell power at the same time.

$$P_{gridt+} + \sum_{g \in S_G} P_{gt} + \sum_{i \in S_I} P^t_{Li} + \sum_{s \in S_S} P_{st-} = P_{gridt-} +$$
$$\sum_{i \in S_L} P^t_{Li} + \sum_{s \in S_S} P_{st+} \qquad (4)$$

$$E_{st} = E_{s(t-1)} - \Delta T(P_{st-} - P_{st+})$$
$$(s \in S_S, t \in S_T) \qquad (5)$$

$$P^{max}_{grid} - P^t_{grid+} + P^t_{grid-} + \sum_{g \in S_G}(P^{max}_{gt} - P_{gt})$$
$$\geq R_{percent}(\sum_{i \in S_L} P^t_{Li}) \quad (t \in S_T) \qquad (6)$$

$$P^{min}_{gt} \leq P_{gt} \leq P^{max}_{gt} \quad (g \in S_G, t \in S_T) \qquad (7)$$

$$\Delta T \cdot P^{Ramp}_g \leq P_{g(t+1)} - P_{gt} \leq \Delta T \cdot P^{Ramp}_g$$
$$(g \in S_G, t \in S_T) \qquad (8)$$

$$U^t_{s+}(P^{min}_S + \Delta P_g) \leq P_{st+} \leq U^t_{s+}(P^{max}_S - \Delta P_g)$$
$$(s \in S_S, t \in S_T) \qquad (9)$$

$$U^t_{s-}(P^{min}_S + \Delta P_g) \leq P_{st-} \leq U^t_{s-}(P^{max}_S - \Delta P_g)$$
$$(s \in S_S, t \in S_T) \qquad (10)$$

$$U^t_{s+} + U^t_{s-} = 1 \quad (s \in S_S, t \in S_T) \qquad (11)$$

$$(E^{min}_s + \Delta E_s) \leq E_{st} \leq (E^{max}_s - \Delta E_s)$$
$$(s \in S_S, t \in S_T) \qquad (12)$$

$$0 \leq P^t_{grid+} \leq U^t_{grid+} P^{max}_{grid} \quad (t \in S_T) \qquad (13)$$

$$0 \leq P^t_{grid-} \leq U^t_{grid-} P^{max}_{grid} \quad (t \in S_T) \qquad (14)$$

$$U_{grid+}^t + U_{grid-}^t \leq 1 \quad (t \in S_T) \tag{15}$$

*B. Real Time Control*

In the RTC phase, the forecasting errors of the dispatch phase and the net-load fluctuation are addressed by quickly adjusting the battery energy storage system. All controllable units' on/off status and outputs remain the same as determined in the RTD phase. In the RTC phase, the power balance equation (4) is solved with actual load and DER generation to determine the battery's charging/discharging rate so that the exchange power with the grid can remain the same as is determined in the RTD phase unless one or multiple constrains of (16) – (18) are violated. The capacity of ESS that is withheld in the RTD phase is now released for net-load fluctuation mitigation in the RTC phase.

$$P_{St+}^{min} \leq P_{st+} \leq P_{St+}^{max} \tag{16}$$
$$P_{St-}^{min} \leq P_{st-} \leq P_{St-}^{max} \tag{17}$$
$$E_s^{min} \leq E_i^t \leq E_s^{max} \tag{18}$$
$$E_{st} = E_{s(t-1)} - \Delta T_C(P_{st-} - P_{st+}) \tag{19}$$

The resulted ESS charging/discharging power after solving (4) should be within the limits defined in (16) – (17). Similarly, the energy stored in energy storage system that is obtained by solving (19) should also be within the limits as well (18).

If all constraints (16) – (18) are satisfied, then the battery energy storage system can fully mitigate the fluctuation of net-load by itself in the real-time control phase. If any of those three constrains are not satisfied, then the grid will have to take over the responsibility to alleviate the fluctuation. The algorithm flow-chart of the RTC phase is shown in Figure 1. The value of 'count' in Figure 1 represents the number of 4-second time intervals that the ESS is not able to fully mitigate the net-load fluctuation.

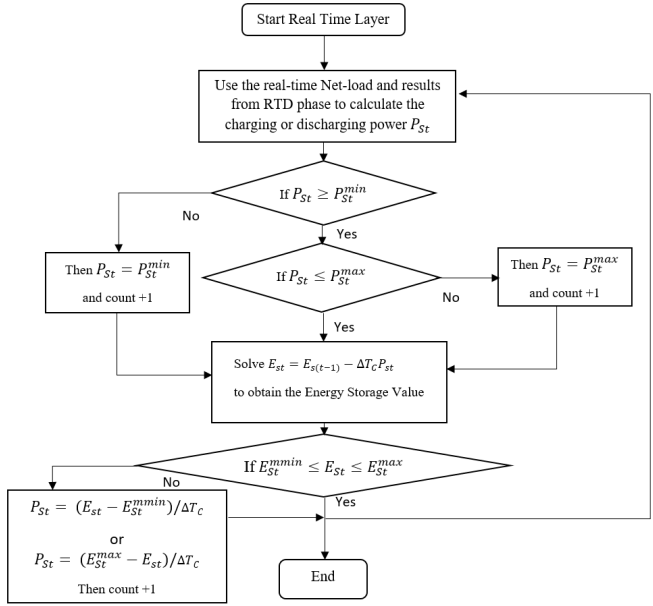

Figure 1. Algorithm flow chart of the RTC phase.

### III. TWO-PHASE REAL-TIME ENERGY MANAGEMENT STRATEGY FOR GRID-FRIENDLY NETWORKED MICROGRIDS

In the RTD phase, all controllable units follow the on/off status obtained from day-ahead scheduling. Model predictive control is applied in this phase to determine the optimal dispatch points. In the RTC phase, the fast-acting battery energy storage system is adjusted to mitigate the sub-minute fluctuation of net-load and to maintain a constant tie-line exchange power between the main grid and microgrid. The proposed two-phase real-time energy management strategy consists of the following steps:

Step 1: Update the forecasting data of load, wind and solar power before the next dispatch run.

Step 2: An economic dispatch problem is formulated to update the dispatch points for each controllable DER unit and energy storage system in the RTD phase. On-off status of controllable units will follow the day-ahead unit commitment solution and not be changed in this step. The RTD phase covers 4 time intervals. The advantage of using MPC in the RTD phase is that it can reduce the negative impact of uncertainties of the forecasting data on the system reliability by looking ahead 4 intervals for each dispatch run. Thus, potential dispatch failure may be avoided. Note that only the solution of first interval will be implemented. MPC is based on moving 'window'. In this paper, the 'window' includes 4 time intervals and each time interval is set to 15 minutes. The moving 'window' has a time length of 1 hour. For example, if a 'window' includes time intervals 1, 2, 3, 4, then an economic dispatch problem is formulated for the 'window' and the optimized dispatch points are solved for intervals 1, 2, 3, 4 while only the dispatch point for interval 1 is implemented. Subsequently, the 'window' will move to the next 4 time intervals (2, 3, 4, 5) after the solution of interval 1 is implemented.

Step 3: The real-time charging/discharging power of ESS in the RTC phase is adjusted every 4 seconds to mitigate real-time net-load fluctuation based on the updated real-time data of load, wind and solar energy. In the RTC phase, each control interval is set to 4 seconds. In order to maintain constant tie-line exchange power flow during an RTD dispatch interval, the ESS adjusts its charging/discharging rate to alleviate the fluctuation of microgrid net-load. When reaching the end of an RTD dispatch interval, the proposed procedure continues by going back to Step 1.

### IV. CASE STUDIES

A typical microgrid system is simulated in this paper. The controllable DER units include a micro-turbine (MT), a fuel cell (FC), and a diesel engine (DE). The load in this simulated microgrid represents 1,000 residential houses. The residential load data are obtained from the Pecan Street Dataport [11]. Non-controllable DER units include solar panels installed in 200 houses (with a capacity of 5kW per house) and four 200kW wind turbine (WT) units. The microgrid also contains an energy storage system with a 500kWh battery set, and its maximum charge/discharge power is 150kW.

In this work, a security-constrained unit commitment (SCUC) is solved by AMPL [12] using Gurobi solver [13] to provide the initial on/off status of controllable DERs, which are fixed in the RTD phase. The time frame under consideration in this paper is 7:00pm─10:00pm. In the RTD phase, each economic dispatch time interval is 15 minutes while the time resolution for the RTC phase is 4 seconds. Assume the net-load does not change in the 4-second RTC interval. The wholesale market price of purchasing power from the main grid is calculated by averaging the hourly price of April, 2020 in Austin, TX. The wholesale market price data were obtained from ERCOT [14]. The price of selling power to the main grid is 80% of the purchasing price. The wholesale purchasing price from 7pm-10pm is shown in Figure 2.

Numerical simulations are conducted on ten different net-load scenarios, of which the forecasting errors are simulated between 1% and 20%. The simulated load, wind turbine generation and solar panel generation are shown in Figure 3−5 respectively. The net-load shown in Figure 6 is the difference between load and variable renewable (wind/solar) generation. Note that Figures 3−8 represent the data and results for a net-load profile with a 5% forecasting error.

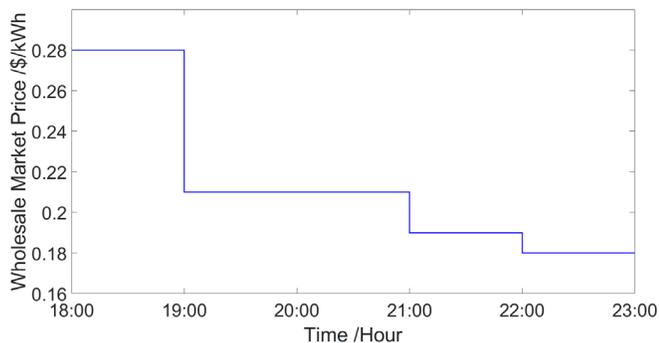

Figure 2. Wholesale market purchasing price.

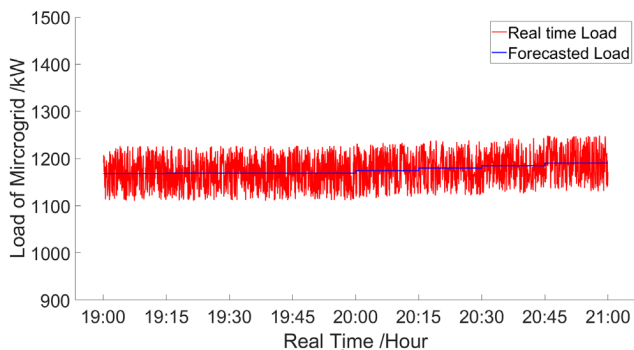

Figure 3. Load of microgrid.

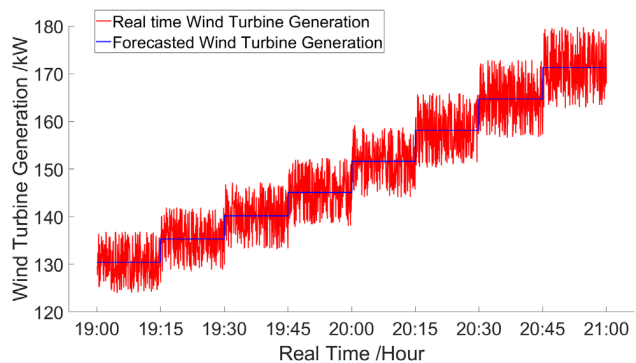

Figure 4. Wind turbine generation of microgrid.

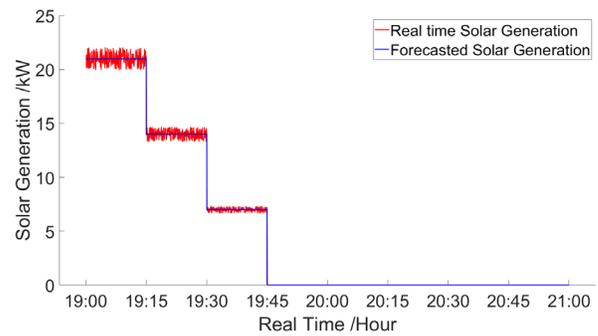

Figure 5. Solar power generation.

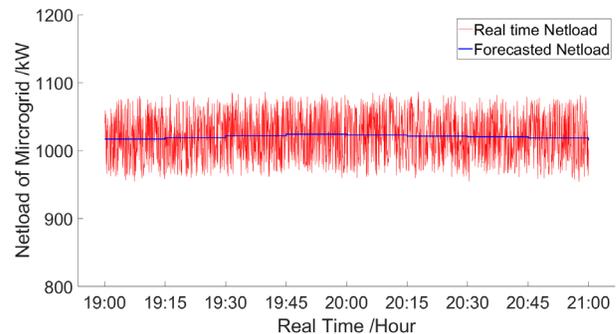

Figure 6. Net-load of microgrid.

The battery charging-discharging power in the RTD phase and the RTC phase is shown in Figure 7. Blue curve denotes the dispatched ESS output power by the RTD phase while red curve denotes the actual ESS output power in the RTC phase. The battery charging-discharging power keeps adjusting in the RTC phase in order to mitigate the fluctuation of real-time net-load.

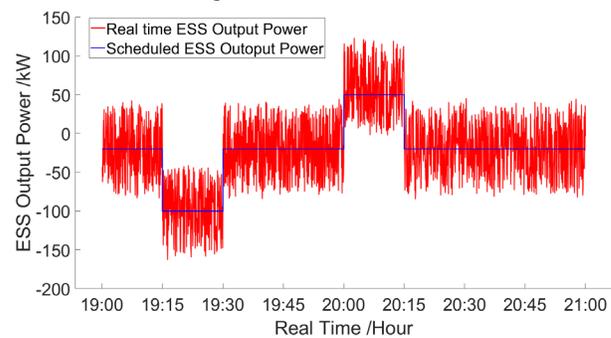

Figure 7. ESS output power from 7pm to 9pm at 5% error.

The tie-line exchange power between the main grid and microgrid is shown in Figure 8. The red curve denotes tie-line exchange power without the RTC while the blue curve denotes the tie-line exchange power with the RTC. After implementing the two-phase energy management strategy, the tie-line exchange power remains constant for each dispatch time interval which is 15 minutes in this work. The real-time net-load fluctuation with a 5% forecasting error are fully mitigated by the battery energy storage system.

With a 5% net-load forecasting error, the proposed strategy can mitigate the net-load fluctuation. Figure 9 shows the tie-line exchange power between the main grid and microgrid when the simulated net-load forecasting error is 10%. It is observed that the tie-line exchange power is not always constant in the second dispatch interval. This is because the magnitude of the net-load fluctuation is very high and requires a power/energy compensation beyond the battery's available capacity. Though this may be addressed by withholding extra power/energy capacity in RTD as shown in (9)−(11), it would affect microgrid operational efficiency as large forecasting error is very rare for a short-term dispatch.

Without the ESS mitigation in the RTC phase, the grid would automatically respond to the microgrid internal net-load fluctuation to maintain microgrid power balance by adjusting the tie-line exchange power, which introduces additional uncertainties into the bulk power grid.

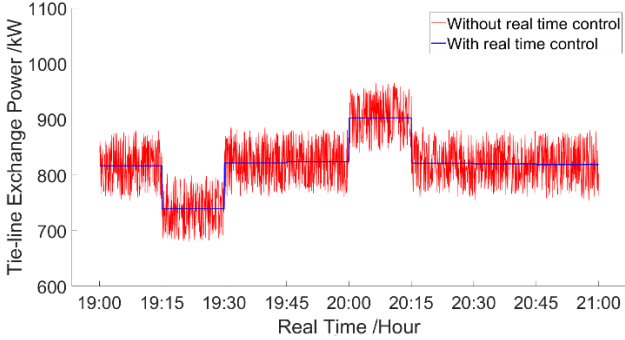

Figure 8. Tie-line exchange power from 7pm to 9pm at 5% error.

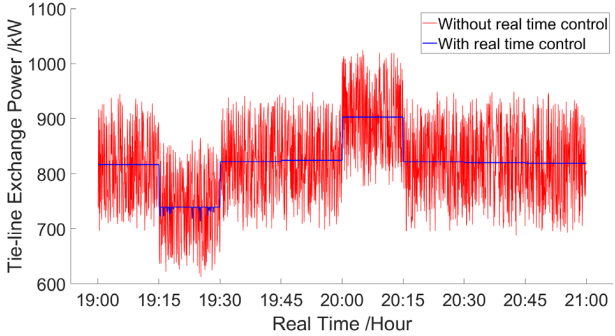

Figure 9. Tie-line exchange power from 7pm to 9pm at 10% error.

A novel metric, Fluctuation Mitigation Rate (FMR), is proposed in this paper to quantify the effectiveness of the proposed two-phase grid-friendly energy management strategy as compared to a regular energy management strategy that consists of a single economic dispatch layer without consideration of real-time net-load fluctuation mitigation. As defined in (20), the proposed metric FMR is a percentage number that represents how much time the microgrid net-load fluctuation is 100% mitigated internally.

$$FMR = \frac{T_1}{T_2} \times 100\% \quad (20)$$

where $T_1$ represents the cumulative time duration when tie-line exchange power keeps at the same level as dispatched in an RTD phase and $T_2$ represents the total time duration of an RTD phase.

For the proposed two-phase energy management strategy, the variance of tie-line exchange power, $v^{PGrid}$ as defined in (21), is much smaller than the traditional single-phase RTD optimization. $v^{PGrid}$ can be used as another metric to quantify the effectiveness of the proposed strategy.

$$v^{PGrid} = \frac{\sum \sigma^2(P_{grid}^c)}{n_D} \quad (21)$$

where $P_{grid}^c$ represents the tie-line exchange power in each 4-second RTC interval $c$. $n_D$ denotes the number of 15-minute dispatch intervals in an RTD phase. $\sigma^2(x)$ denotes the variance of variable $x$.

Table I shows the grid exchange power statistics over results associated with 10 different net-load scenarios. When the forecasting error is between 0 to 5%, the variance of tie-line exchange power is 0 and the FMR is 100% for the proposed two-layer approach while the counterpart variance for a traditional approach without the real-time net-load fluctuation mitigation control phase goes up to over 1,000. As the forecasting error increases, the variance of tie-line exchange power deviation for the proposed method also increases. This leads to the decrease of the FMR which means the tie-line exchange power needs to be changed to balance the microgrid. Note that very large forecasting error is not common for short-term operations.

Table I. Statistics of real-time tie-line exchange power for 10 different scenarios corresponding to 10 different forecasting errors

| Error | without RTC | | with RTC | |
|---|---|---|---|---|
| | $v^{PGrid}$ | FMR | $v^{PGrid}$ | FMR |
| 0.5% | 11.46 | 0% | 0 | 100% |
| 1% | 45.29 | 0% | 0 | 100% |
| 2% | 183.86 | 0% | 0 | 100% |
| 3% | 407 | 0% | 0 | 100% |
| 4% | 741 | 0% | 0 | 100% |
| 5% | 1191 | 0% | 0 | 100% |
| 8% | 2492 | 0% | 0.0023 | 99.89% |
| 10% | 4599 | 0% | 1.4491 | 99.11% |
| 15% | 10722 | 0% | 1757 | 95.11% |
| 20% | 19269 | 0% | 9485 | 82.72% |

Table II shows tie-line exchange power statistics for a net-load scenario under a forecasting error of 10%. If the minimum and maximum values of tie-line power flow in a dispatch interval are the same, then the ESS can fully mitigate the real-time net-load fluctuation. This holds for all intervals except for the second interval. For the second time interval (19:15-19:30), $v^{PGrid}$ with RTC is not zero, which indicates that the second interval does not keep a constant tie-line exchange power, which is consistent with the results shown in Figure 9. The $v^{PGrid}$ without RTC is much higher, which indicates that the microgrid net-load fluctuation if not addressed internally can create significant uncertainty to the bulk grid.

Table II. Statistics of real-time tie-line exchange power for a net-load scenario under a forecasting error of 10%

| Time | without RTC | with RTC | | | |
|---|---|---|---|---|---|
| | $v^{PGrid}$ | $v^{PGrid}$ | Tie-line flow (kW) | | FMR |
| | | | Min | Max | |
| 19:00 -19:15 | 4619 | 0 | 817 | 817 | 100% |
| 19:15-19:30 | 4505 | 11.6 | 713 | 739 | 92.89% |
| 19:30-19:45 | 4356 | 0 | 822 | 822 | 100% |
| 19:45-20:00 | 4218 | 0 | 824 | 824 | 100% |
| 20:00-20:15 | 5015 | 0 | 903 | 903 | 100% |
| 20:15-20:30 | 4428 | 0 | 822 | 822 | 100% |
| 20:30-20:45 | 4877 | 0 | 820 | 820 | 100% |
| 20:45-21:00 | 4733 | 0 | 819 | 819 | 100% |

## V. Conclusion

A novel two-phase real-time energy management strategy for grid connected microgrids is proposed in the paper. It consists of an RTD phase and an RTC phase. In the RTD phase, MPC is applied to determine the microgrid dispatch points. In the RTC phase, the energy storage system is used to mitigate the fluctuation in order to maintain a constant tie-line exchange power between the main grid and microgrid for each dispatch interval. As shown in the simulation results, the proposed two-phase real-time energy management strategy can effectively decrease the variance of grid exchange power for a microgrid dispatch interval and mitigate the real-time sub-minute net-load fluctuation in a microgrid. In other words, the proposed strategy can harness the microgrid internal fluctuation internally and thus, relieve the negative impact of the uncertainties of microgrid net-load and contribute to enhancing the reliability of the bulk power grid. To summarize, with the proposed two-phase energy management strategy, a microgrid can be considered as a grid-friendly microgrid from the perspective of a bulk grid operator.